\documentclass[journal]{IEEEtran}
\ifCLASSINFOpdf
   \usepackage[pdftex]{graphicx} 
\else
   \usepackage[dvips]{graphicx}
  \graphicspath{{../eps/}}
\fi
\usepackage[cmex10]{amsmath}
\usepackage{caption}
\usepackage{float}
\usepackage[table]{xcolor}
\usepackage{multirow}
\usepackage{graphics}

\usepackage{multicol}
\usepackage{cuted}
\usepackage{enumerate}
\usepackage{subfigure}
\usepackage{acronym}
\usepackage{epstopdf}
\usepackage{cite}
\usepackage{xcolor}
\usepackage{fancyhdr}
\usepackage{amsfonts}
\usepackage{titlesec}
\usepackage{tikz}

\title{Revisiting the Wireless Channel from Physical Layer Security Perspective}
	\author{Abuu. B. Kihero,~Haji. M. Furqan, ~M. M. \c{S}ahin,~and~H{\"u}seyin~Arslan,~\IEEEmembership{Fellow,~IEEE}%
	\thanks{A. B. Kihero,~H. M. Furqan, and H. Arslan are with the Department of Electrical and Electronics Engineering, Istanbul Medipol University, Istanbul, 34810 Turkey. H. Arslan and M. M. Sahin with Department of Electrical Engineering, University of South Florida, Tampa, FL 33620 USA.}
	}

\makeatother
\acrodef{PLS}[PLS]{physical layer security}
\acrodef{OFDM}[OFDM]{orthogonal frequency domain multiplexing}
\acrodef{MAC}[MAC]{Medium Access Control}
\acrodef{ISI}[ISI]{inter-symbol interference}
\acrodef{RF}[RF]{radio frequency}
\acrodef{CF}[CF]{carrier frequency}
\acrodef{EM}[EM]{electromagnetic}
\acrodef{AoA}[AoA]{angle of arrival}
\acrodef{ADoA}[ADoA]{angle difference of arrival}
\acrodef{AoD}[AoD]{angle of departure}
\acrodef{ToA}[ToA]{time of arrival}
\acrodef{TDoA}[TDoA]{time difference of arrival}
\acrodef{LoS}[LoS]{line-of-sight}
\acrodef{NLoS}[NLoS]{non-\ac{LoS}}
\acrodef{RSS}[RSS]{received signal strength}
\acrodef{RSSI}[RSSI]{\ac{RSS} indicator}
\acrodef{CSI}[CSI]{channel state information}
\acrodef{CIR}[CIR]{channel impulse response}
\acrodef{CFR}[CFR]{channel frequency response}
\acrodef{rtt}[RTT]{round trip time}
\acrodef{PL}[PL]{path loss}
\acrodef{MPC}[MPC]{multipath component}
\acrodef{mmWave}[mmWave]{millimeter wave}
\acrodef{THz}[THz]{Terahertz}
\acrodef{WPE}[WPE]{wireless propagation environment}
\acrodef{V2V}[V2V]{vehicle-to-vehicle}
\acrodef{HST}[HST]{high speed train}
\acrodef{V2X}[]{short= V2X, long= vehicle-to-everything}
\acrodef{MIMO}[]{short= MIMO, long= multiple input multiple output}
\acrodef{mMIMO}[mMIMO]{massive multiple input multiple output}
\acrodef{ULA}[ULA]{uniform linear array}
\acrodef{RA}[RA]{reconfigurable antenna}
\acrodef{MBM}[MBM]{media-based modulation}
\acrodef{SNR}[SNR]{signal-to-noise ratio}
\acrodef{WSSUS}[WSSUS]{wide sense stationary uncorrelated scattering}
\acrodef{WSS}[WSS]{wide sense stationary}
\acrodef{VR}[VR]{visibility region}
\acrodef{SW}[SW]{spherical wavefront}
\acrodef{UWB}[UWB]{ultrawide band}
\acrodef{ML}[ML]{machine learning}
\acrodef{UE}[UE]{user equipment}
\acrodef{Wi-Fi}[Wi-Fi]{wireless fidelity}
\acrodef{lpi}[LPI]{low probability of intercept}
\acrodef{JRC}[JRC]{joint radar-communication}
\acrodef{IoT}[IoT]{internet-of-things}
\acrodef{SKG}[SKG]{security key generation}
\acrodef{vlc}[VLC]{visible light communication}
\acrodef{gps}[GPS]{global positioning system}
\acrodef{gnss}[GNSS]{global-navigation satellites system}
\acrodef{uav}[UAV]{unmanned aerial vehicle}
\acrodef{lidar}[LiDAr]{Light Detection and Ranging}
\acrodef{RTI}[RTI]{radio tomographic imaging}
\acrodef{MA}[MA]{multiple access}
\acrodef{PDMA}[PDMA]{path division multiple access}
\acrodef{VA}[VA]{virtual anchor}
\acrodef{TDD}[TDD]{time division duplex}
\acrodef{FDD}[FDD]{frequency division duplex}
\acrodef{SINR}[SINR]{signal-to-interference-plus-noise ratio}
\acrodef{BER}[BER]{bit error rate}
\acrodef{ROC}[ROC]{receiver operating characteristic}
\acrodef{RIS}[RIS]{reconfigurable intelligent surface}
\acrodef{5G}[5G]{fifth generation}
\acrodef{6G}[6G]{sixth generation}
\acrodef{BS}[BS]{base station}
\acrodef{MoA}[MoA]{molecular absorption}
\acrodef{MoS}[MoS]{molecular scattering}
\acrodef{NTN}[NTN]{non-terrestrial network}
\acrodef{ABC}[ABC]{ambient backscatter communication}
\acrodef{REM}[REM]{radio environment mapping}
\acrodef{DD}[DD]{delay diversity}
\acrodef{CDD}[CDD]{cyclic \ac{DD}}
\acrodef{DoD}[DoD]{Doppler delay diversity}
\acrodef{FIR}[FIR]{finite impulse response}
\begin{document}
\maketitle

\begin{abstract}
Security has emerged as one of the critical requirements in future wireless networks. Unlike traditional cryptography-based security, physical layer security (PLS) exploits various features of the random wireless channel to secure not only the information being communicated but the whole communication process from any type of attack. Future wireless networks are envisioned to feature new technologies such as re-configurable intelligent surfaces, massive multiple input multiple output, and sensing, to accommodate the emerging use-cases. Both, the new technologies and the new use-cases have been found to enrich the channel characteristics by unveiling some new channel features which can be readily exploited to facilitate PLS. This article surveys these new channel features to reveal their potential for PLS implementation. In the course of the article, the assessments of important qualities while selecting a certain channel feature for the PLS application are discussed. The importance of the channel control concept and sensing technologies that facilitate the accessibility of certain channel features are highlighted from the PLS perspective. Security attacks aimed at channel characteristics, rather than the communication itself, in order to disrupt PLS implementations are also discussed. Finally, the article summarizes the possible research direction for channel-based PLS.
\end{abstract}
\IEEEpeerreviewmaketitle
\section{Introduction}
The scope of wireless technology has been constantly growing and its focus has shifted from voice calls only in the earliest days to the internet of everything in the most recent generations. Such widening scope has been driven by the incessant growth of the societal needs that have resulted in the emergence of myriads of use cases and scenarios with diversified sets of performance requirements. Specifically, spectral and energy efficiency, reliability, latency, seamless connectivity, and support for high mobility along with the traditional quest for a higher data rate have emerged as critical requirements. Above all these, considering the broadcast nature of the wireless signal, the security of the communication process has become exceedingly crucial for most, if not all of the prospective use cases and scenarios.

Different from the traditional security concern that mainly considers the privacy of the information being communicated, in the recent networks the physical signal (its waveform, beam, energy, path, etc.), radio resources (spectrum, power, etc.), user information (location, context, health, etc.), propagation channel and the radio environment as whole are prone to numerous types of attacks such as eavesdropping, manipulation, spoofing, and jamming. Consequently, the traditional cryptographic approaches are not enough to provide the desired level of security. To this end, \ac{PLS} emerged as a promising and revolutionizing concept to address these security concerns \cite{hamamreh2018classifications}.

\begin{figure}[t]
    \centering
\includegraphics[width=1\columnwidth]{ 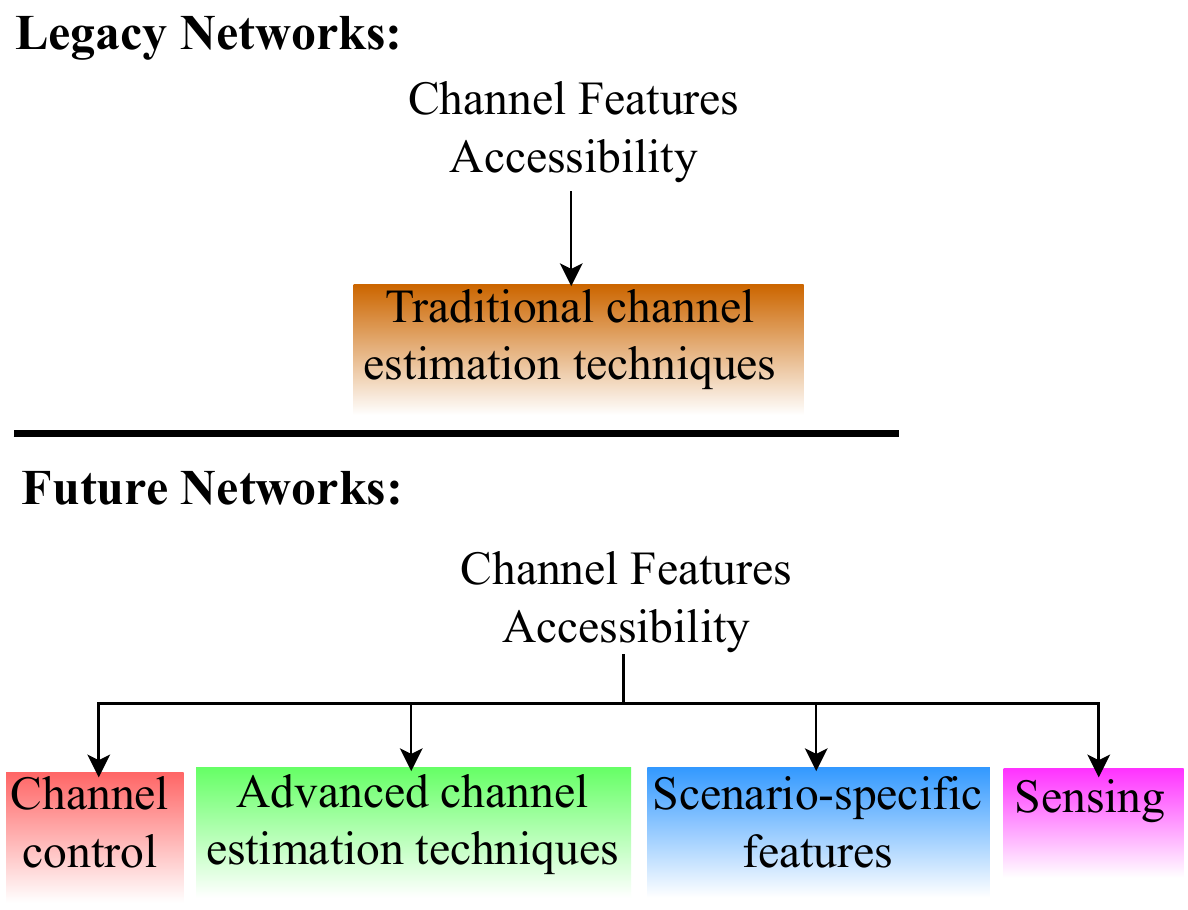}
    \caption{\footnotesize Comparison of the legacy and future networks' abilities in interacting with the propagation conditions}
    \label{fig:featuresAccessibility}
\end{figure}

\ac{PLS} is based on exploiting the dynamic characteristics of the wireless environment including the propagation channel, \ac{RF} front-end, and/or the communication signal itself to secure the communication process without needing to share the security keys. 
While there are multiple dimensions of the communication process that can be leveraged for \ac{PLS} as referred above, this article focuses on the popular, wireless-channel based \ac{PLS} implementation. Undoubtedly, propagation channel characteristics have been extensively exploited for \ac{PLS}. However, only limited features of the channel, such as \ac{RSS}, amplitudes, and/or phases, have been considered by most of the \ac{PLS} techniques. 
This is due of the nature of legacy networks which did not give much accessibility to some other channel features. The recent advancement in transceiver architectures, signal processing, and channel estimation techniques, numerous channel features have come into prominence and made available for exploitation, individually or jointly. The recent advancement in technology have made numerous channel features available for exploitation, individually or jointly. Additionally, next-generation networks are also expected to feature sensing as well as channel control mechanisms which will not only improve the quality of the acquired \ac{CSI} but also enable the network to optimize the propagation characteristics based on the ongoing communication. Figure \ref{fig:featuresAccessibility} compares legacy and future networks in terms of their capabilities in interacting or gaining awareness of the propagation condition. Although significant effort has been invested in exploiting wireless channel for \ac{PLS} \cite{hamamreh2018classifications}, no study has been dedicated to give deep insight on the individual channel characteristics from \ac{PLS} perspective in the light of the newly acquired network capabilities. Therefore, this study sets to explore \ac{PLS} potential of the nascent channel characteristics enabled by these new network capabilities. To this end, this article touches the following wireless channel aspects from \ac{PLS} perspective:

\begin{itemize}
\item \textit{Identifying the eligibility criteria that make a given channel feature suitable for \ac{PLS}:} Even though it will be possible to observe and access different channel features in the future networks, not every feature is suitable for \ac{PLS} application. Essentially, the suitability of a given feature strictly depends on the nature of the attack against which a \ac{PLS} technique is devised. Traditionally, randomness, reciprocity, and spatial decorrelation of the channel response are considered to be the main requirements desired for a successful \ac{PLS} realization. However, in the literature, scopes of these criteria are constrained to specific types of attacks, particularly, eavesdropping attacks. Considering the innumerable types of security threats that have emerged, this article intends to emphasize on the wider understanding of such eligibility criteria.

\item \textit{Exploring the potential of various channel features on \ac{PLS}:} Various measurement and numerical analyses pertaining to new use cases and network features, such as \ac{mMIMO}, \ac{V2X}, \ac{UWB}, etc., have revealed numerous interesting channel characteristics that are not observable in the legacy systems. A number of studies on how to exploit these characteristics to improve different aspects of communication systems, such as capacity and channel estimation, are already underway. This study in particular traverses the \ac{PLS} aspects of these channel characteristics, individually, highlighting the opportunities and exposing potential challenges.

\item \textit{Role of channel control and sensing in \ac{PLS}:}
The next-generation wireless networks are envisioned to feature channel control and sensing capabilities through smart radio environment and \ac{JRC} frameworks, respectively. While sensing gives the extra capability for the network to deeply understand the propagation environment and adapt itself accordingly, channel control allows it to optimize the propagation characteristics in favor of the ongoing communication. This article will explore how these technologies can be instrumental in improving \ac{PLS}.

\item \textit{Integrity of the channel features for \ac{PLS}:} Finally, the study highlights attacks dedicated directly to the channel characteristics, making them unsuitable for \ac{PLS} applications. That is, some attacks focus on disrupting the desirable channel features such as reciprocity and spatial decorrelation such that the channel becomes inappropriate for \ac{PLS} applications.
\end{itemize}
Although significant effort has been invested in exploiting wireless channel for \ac{PLS}, no study has been dedicated to give deep insight on the individual channel characteristics from \ac{PLS} perspective in the light of the newly acquired network capabilities.

\section{Eligibility Requirements of Channel Parameters for PLS}
In this section, typical requirements that make a given channel feature suitable for \ac{PLS} are discussed, where
Fig. \ref{fig:ChannelFundamentalParameters} summarized some minimum criteria for any feature to be considered exploitable for \ac{PLS} applications.
In order to pave the way for better comprehension of those requirements, a quick recap on the basic wireless channel-based \ac{PLS} approaches is given along with the types of attacks that they are able to address.

\subsection{Channel-based \ac{PLS} Approaches}
\subsubsection{Channel-based Key Generation}
Unlike the traditional Diffie-Hellman key-exchange mechanism in which the level of communication security relies on the computational complexity of the cryptographic keys, the channel-based \ac{SKG} avoids the key sharing process by employing the wireless channel characteristics as a common source of randomness between legitimate nodes to generate the inherently shared keys. The success of this approach relies mainly on the reciprocity and spatial/temporal decorrelation of the observed channel response such that the generated key is unique to the channel of legitimate nodes. The channel-based \ac{SKG} approach is mainly used to facilitate confidentiality against eavesdropping attacks or as a means of authentication against spoofing attacks.

Generally, the key is generated by quantizing the observed channel response to create the secret binary bits. The success of this approach relies mainly on the reciprocity and spatial/temporal decorrelation of the observed channel response such that the generated key is unique to the channel of legitimate nodes. The channel-based \ac{SKG} approach is mainly used to facilitate confidentiality against eavesdropping attacks. However, the generated key can also be employed as a means of authentication against spoofing attacks. In some cases, the \ac{SKG} approach is also used to generate the channel-based spreading codes to provide robustness against jamming attacks.

\subsubsection{Channel-based Adaptation \ac{PLS} Techniques}
Similar to the \ac{SKG}, this approach also makes use of the channel response as a shared source of randomness to secure communication. But, instead of generating secrecy keys, legitimate nodes dynamically adapt their transmission parameters based on the instantaneous channel condition which is unique to them. This approach is applicable against wide range of attacks, including eavesdropping, spoofing, jamming, impersonation, etc.  Examples include beamforming, pre-coding, cooperative relaying, \ac{RIS} based adaptation, and so on.

\subsubsection{Interference-based Techniques}
Here, null-space of the channel between the legitimate nodes is exploited by transmitting interference signals to degrade decoding performance at the attacker's node. Interference signals are generally in the form of artificial noise or friendly jamming signals. Relying on the spatial decorrelation of the fading characteristics assumption, null-spaces of the legitimate and illegitimate node's channels are different and thus the transmitted interference signal is ought to hurt the illegitimate nodes only. This approach has been mainly employed against eavesdropping and spoofing attacks.

\begin{figure}[t]
    \centering
\includegraphics[width=0.8\columnwidth]{ 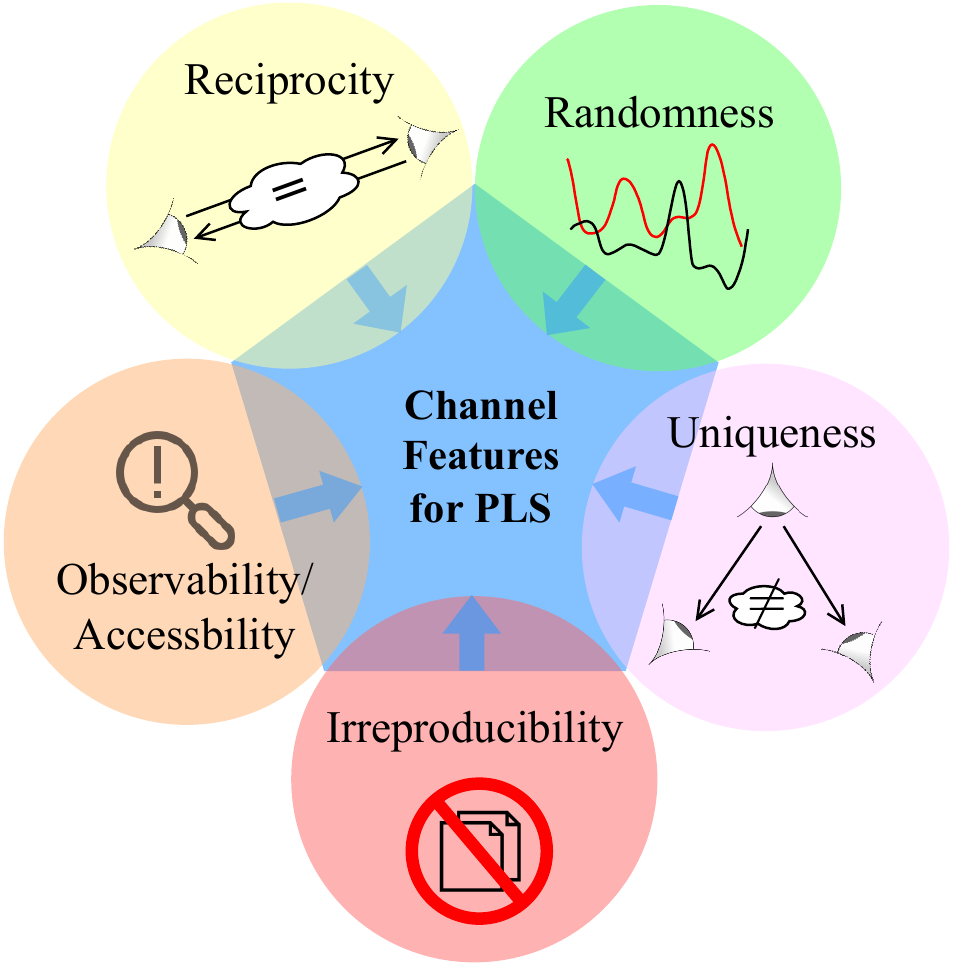}
    \caption{\footnotesize Desirable qualities of a channel feature suitable for \ac{PLS} application.}
    \label{fig:ChannelFundamentalParameters}
\end{figure}

\subsection{Eligibility Criteria}
\Ac{PLS} techniques can utilize different features of the wireless channel to secure the wireless signals and the contents carried therein. For a certain channel feature to be considered exploitable for \ac{PLS} applications, it has to fulfill some minimum criteria which are summarized in Fig. \ref{fig:ChannelFundamentalParameters} and explained below. It is important to emphasize that the importance of a given criterion highly depends on the type of attack against which a \ac{PLS} technique is devised.

\subsubsection{Randomness}
The random variability of the channel characteristics in different domains is an essential requirement as it increases channel entropy for stronger \ac{PLS} implementation. Generally, the randomness of the channel response is mainly due to the characteristics of the environment scatterers as well as the communicating terminals. Obviously, static and deterministic channel characteristics are unsuitable to exploit for \ac{PLS} applications as they can be easily predicted by the attackers. Essentially, the randomness of the channel response in a given domain is defined by the channel selectivity in that domain. Accordingly, channel's coherence parameters, i.e.,  coherence-time/bandwidth/space, are generally used to measure channel randomness in a respective domain. From \ac{PLS} perspective, randomness is important specifically for the channel-based \ac{SKG} and adaptation security approaches. For example, in the case of the channel-based \ac{SKG} approaches, the level of randomness in the channel determines the entropy of the generated key where high entropy leads to stronger keys. In a typical \ac{SKG} process that extracts the key from the fading pattern of the channel, the measured channel response is quantized to generate the secrecy bits. Therefore, the higher the randomness the more the quantization levels and the longer the secrecy bits stream. Moreover, the highly random channel characteristic is more difficult to reproduce or mimic as in the case of the impersonation attacks.


\subsubsection{Uniqueness}
This quality insists that the channel response observed by the legitimate node is exclusive to them only with respect to their instantaneous context such that whatever security parameters extracted therefrom cannot be replicated anywhere else or at any other time instant. This is basically the continuation of the randomness requirement where by the channel dynamicity and spatial decorrelation are emphasized. The channel, apart from being selective/random in the domain of interest within a single channel usage, it has to be dynamic such that the values of secrecy parameters derived at one instant are not the same with the ones at a later time. This is strictly important especially when exploiting frequency or spatial randomness of the channel. For instance, a channel can be highly selective in frequency domain in the presence of remote scatterers (i.e., longer delay spread), however, if the scenario is completely static the same channel response is observed in each channel usage. This may tempt the adversary to run her own measurement later and estimate the used security key.  It is also important to point out that rapid variation of the channel may lead to the channel aging problem which can completely disrupt not only security but the whole communication performance in general. 
Apart from the temporal uniqueness, the spatial uniqueness, which mainly referred to as \textit{spatial decorrelation}, of the channel is also necessary. This quality insists that the channel response observed by the legitimate node is exclusive to them only with respect to their instantaneous context such that whatever security parameters extracted therefrom cannot be replicated anywhere else. In general, the uniqueness or spatial decorrelation of a channel is determined by its coherence distance. In theory, in rich scattering propagation conditions, the uniqueness of the channel of legitimate nodes is assumed to be maintained when the attacker's node is at least a half-wavelength away. However, considering the sparse nature of the channels observed in the recent networks due to the usage of \ac{UWB} signaling (temporal sparsity), \ac{mMIMO} (spatial sparsity), a high-frequency band (spatial and temporal sparsity), etc., this assumption might be wanting. Abiding by this requirement has been extremely challenging especially when the attacker is co-located with one of the legitimate nodes. Therefore, channel parameters that have a high degree of spatial independence are more appealing for security.
\subsubsection{Reciprocity}
Although a suitable channel feature is the one that varies randomly and continuously, measurements of that feature at the legitimate nodes must be highly correlated at any given instant. This is in order to ensure that the considered feature serves as the common (inherently shared) source of randomness between the legitimate nodes, which is the prime motivation of the channel-based \ac{PLS} approaches. Imperfect reciprocity may not only compromise the security level but also degrade communication performance. In the \ac{SKG} techniques, the information reconciliation step is usually employed to correct any incurred key mismatch due to imperfect reciprocity which can lead to the leakage of some of the secret bits and jeopardize the confidentiality. Considering the temporal dynamicity of the channel discussed above, in order to preserve reciprocity, it is necessary that the legitimate nodes complete the bidirectional channel probing for \ac{CSI} acquisition before the channel decorrelates over time. However, if this dynamicity is too rapid, the channel aging problem may also occur where the channel response observed during probing stage is different from the one observed during the transmission. This can completely disrupt not only the security but the communication in general. 

\subsubsection{Accessibility/Observability}
This refers to the amount of processing required to access a given channel feature and utilize it for \ac{PLS}. Considering the exponential increase in the number of the low-end, low-energy, and lightweight computing \ac{IoT} devices whose needs for security are in no way less important, easy accessibility of the channel features for \ac{PLS} is of paramount importance. For example, although \ac{mMIMO} inherently provides spatial resolution of the channel, an extra virtual/beamspace channel processing is required to extract (and thus utilize) the \acp{AoD}/\acp{AoA} of the \acp{MPC}. Apart from easiness, the quality with which the feature is accessed is also important. This considers the robustness of the feature against estimation errors, noise, or other \ac{RF} front-end impairments. In the \ac{SKG} approach, for instance, noise can devastatingly decrease the secrecy bits agreement ratio. 

\subsubsection{Irreproducibility}
Although the \textit{uniqueness} requirement mentioned above ensures that the legitimate node and the attacker observe different channels, it does not take away the liberty of the attacker to emulate features of the legitimate nodes' channel. The attacker can manipulate her own channel by injecting fake wireless channel characteristics and making it appear as a legitimate channel response to the target receiver. This is referred to as camouflage attack and is prevalent in wireless channel-based localization applications \cite{fang2014you}. It is a form of impersonation attack. Therefore, the channel features that are not only unique but also difficult to emulate are more preferred against these kinds of attacks.

\begin{figure*}[t]
  \centering
\includegraphics[width=1.95\columnwidth]{ 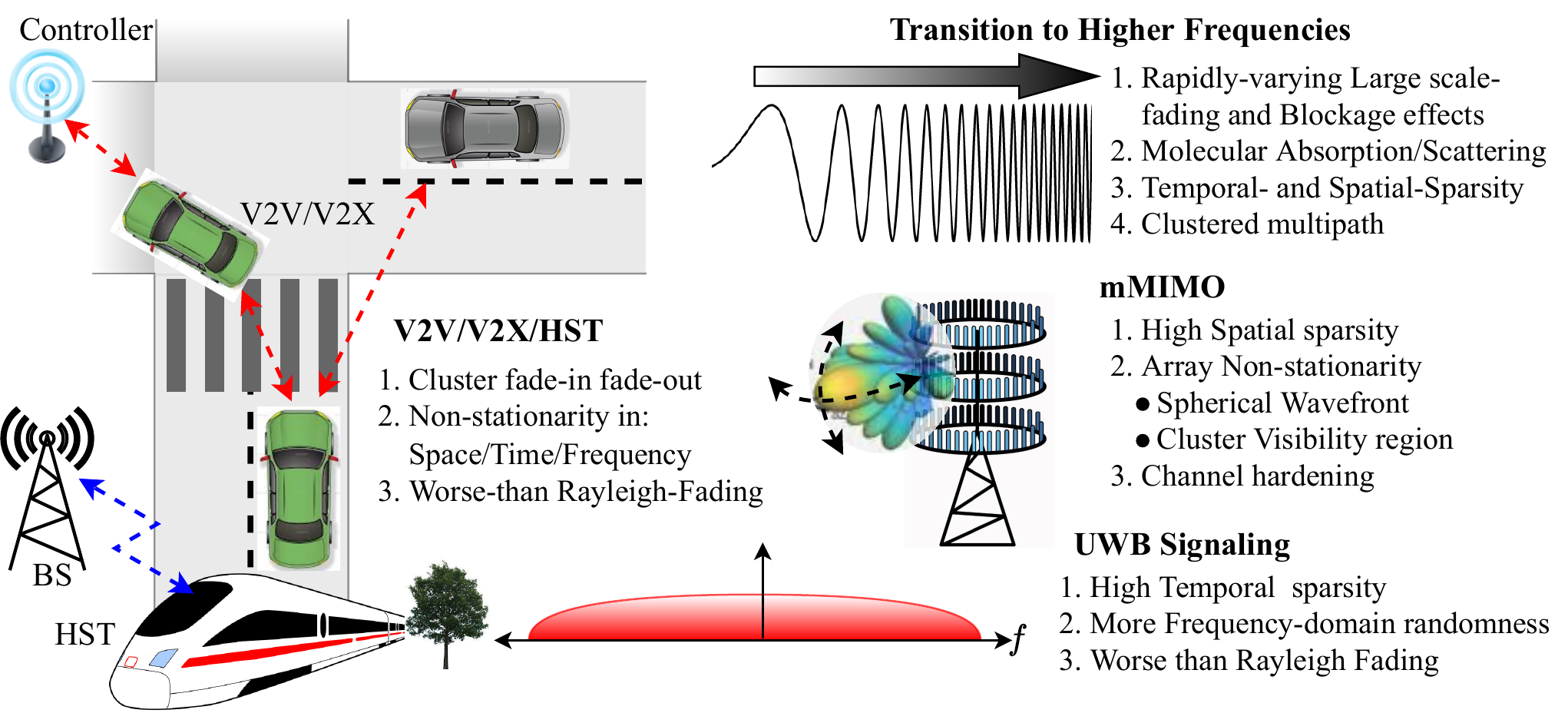}
\caption{\footnotesize Unprecedented channel features due to the use cases and technologies featured in 5G and beyond wireless networks.}
\label{fig:ChannelFeatures}
\end{figure*}

\section{Channel Parameters Beyond 5G: PLS Perspective}\label{sec:ChannelParameters}
This section explores the new channel features that have emerged or become prominent due to the inclusion of some technologies and use cases (summarized in Fig. \ref{fig:ChannelFeatures}) in the wireless networks from \ac{PLS} perspective.  It is however important to emphasize that some of these features are still being studied and only their tentative models are available. Therefore, the discussion herein only intends to reveal their potential from the \ac{PLS} point of view and encourage further research on that direction. 

 Until now, only limited features of the channel have been significantly exploited for \ac{PLS}, namely \ac{RSS}, channel gains, envelope, and phase. This has been highly contributed by the nature of the legacy networks which did not give much accessibility to other features/parameters of the channel. Considering the recent advancement in the transceiver architectures, signaling techniques, signal processing, and channel estimation techniques, numerous channel features have come into prominence and made available for exploitation, individually or jointly. Some of these features are summarized in Fig. \ref{fig:ChannelFeatures}. This section aims at shedding some light on these new channel features from \ac{PLS} perspective. It is however important to emphasize that some of these features are still being studied and only their tentative models are available. As such, the discussion herein only intends to reveal their potentiality from \ac{PLS} point of view and encourage further research on that direction. 

\subsection{Large-scale Fading}
The deterministic nature of the distance-dependent \ac{PL} and the slow-varying shadowing phenomenon have always rendered the large-scale fading aspect of the channel less attractive for \ac{PLS}. However, unlike in the legacy communication networks that mostly operate on microwave frequencies where large-scale fading varies extremely slowly, the recent shift to higher frequency bands like \ac{mmWave} in 5G and the prospected \ac{THz} in 6G and beyond has changed this paradigm. Owing to the small wavelengths at these higher frequencies, even small objects in the environment can cause significant momentary shadowing or even total blockage of the signal, leading to the rapidly and randomly varying signal strength across space and/or time. This increases the level of randomness in the channel and reduces channel correlation in space and time, thereby facilitating stronger key-based and adaptation-based \ac{PLS} techniques. It is also revealed in \cite{karttunen2017spatially} that, at higher frequencies, \ac{PL} coefficients are significantly affected by streets and buildings orientations, leading to the significant variation of the signal strength between close by streets at the same distance from the \ac{BS}. These nascent behaviors of PL and shadowing introduce another layer of randomness into the channel which signifies the potential of large-scale fading in facilitating \ac{PLS}. Additionally, high \ac{PL} effect observed at higher frequency bands bestows a high degree of spatial confinement of the communication signal that provides inherent security. Some studies such as \cite{sarkar2012secrecy} has already shown that, with some proper power allocation schemes and transmission strategies, large-scale fading can be individually exploited to provide security against eavesdropping attacks. Essentially, considering the fact that small and large scale fading phenomena occur concurrently, the traditional \ac{PLS} approaches that mainly focus on small scale fading characteristics of the channel can be revisited to leverage the rapidly varying \ac{PL} and shadowing to enhance their achievable secrecy level.

\subsection{Molecular Absorption and Scattering}
At higher frequencies, the channel exhibits not only peculiar characteristics related to \ac{PL} and shadowing but also a severe form of frequency-selective absorption known as \ac{MoA}. Some of the frequencies in \ac{mmWave} and \ac{THz} bands align with natural resonance frequencies of the atmospheric contents like Oxygen and water molecules. When excited at their resonance frequencies, these molecules absorb a significant amount of energy from the signal which consequently elevates \ac{PL} at these particular frequencies. However, shortly after, some of this energy is re-radiated to the environment at the same frequency. Although some studies have treated the re-radiated energy as a source of noise in addition to the conventional thermal noise, experiments have revealed that this energy is highly correlated with the original signal and thus should be treated as the scattered copy of the signal (i.e., \ac{MoS}). As such, \ac{MoS} has been found to increase the multipath richness of the channel (lowering the channel's $K-$factor), thereby increasing the randomness which is beneficial for \ac{PLS}. However, this scattering effect can also have undesirable impacts. The scattered signal can be exploited by eavesdroppers to listen to the ongoing communication. Furthermore, in case of atmospheric turbulence, heavy rain/snow, \ac{MoA} and \ac{MoS} can be strong enough to completely destroy the beam and disrupt communication between the legitimate nodes.

\subsection{Small-Scale Fading}
Multipath fading is arguably the most exploited feature of the channel for \ac{PLS}. This is mainly because this feature readily meets most of the afore discussed eligibility conditions for \ac{PLS}. Since multipath fading is a superposition of random replicas of the transmitted signal, it is inherently random. Generally, multipath characteristics are function of the propagation environment in terms of the number, distribution, and mobility of the scatterers with respect to the locations of the communicating terminals, which makes them not only random but also spatially unique.

Reciprocity is another inherent quality of multipath fading. The multipath propagation phenomenon is essentially made up of some frequency-dependent and frequency-invariant instances. At lower frequency bands, the channel is often rich scattering, making it difficult to distinguish and exploit these instances individually. Accordingly, majority of the \ac{PLS} techniques in the legacy generations assume the validity of this reciprocity property in \ac{TDD} systems only. Nevertheless, it is reported in \cite{vasisht2016eliminating} that wireless channels at one frequency band can be inferred from other bands. The basic principle is that, for nearby bands, the physical paths traversed by the signal remain the same even though the overall channel response changes. The authors developed a $H_1-$\textit{physical-paths}$-H_2$ mapping mechanism to estimate channel $H_2$ at one band from a known channel $H_1$ at a different band. Development of such techniques enables \ac{PLS} implementation even in the absence of the true reciprocity like in \ac{FDD} systems where uplink and downlink transmissions use different bands. 

Classically, remote scatterers cause large delay spread which is the primary source of the frequency selectivity, whereas scatterers local to the transceiver terminals are the main source of time selectivity in the presence of mobility. Likewise, the scatter distribution with respect to the communicating terminals determines the spatial selectivity of the channel. Generally, a rich scattering environment with some mobile scatterers/terminals imparts more randomness compared to the poor scattering static environment. Furthermore, the absence of the strong \ac{LoS} component makes the channel more random.

While the randomness of the fading coefficients is a key factor for generating strong security keys, the uniqueness of these keys relies on how unique the coefficients are. The uniqueness of the multipath channels is strictly determined by the channel's coherence parameters. Coherence parameters determine how far apart two-terminal should operate in spatial-, time-, or, frequency-domain to observe uncorrelated (unique) and random fading pattern at the legitimate and illegitimate nodes. Short coherence-time/-bandwidth/-distance are preferable for different aspects of \ac{PLS}, such as solutions against eavesdropping and impersonation attacks. Nevertheless, they can also have undesirable impacts on other aspects of \ac{PLS}. For instance, in the case of channel-based authentication against impersonation attacks, frequent updates of the channel database are to be required when the channel varies rapidly, which complicates the process.

As pointed out before, while the rapidly varying channels are good from randomness and uniqueness perspectives, they cause channel aging problem that distracts the channel reciprocity, degrade the performance of \ac{PLS} algorithms. 
\subsection{Sparsity}
\begin{figure}[t]
    \centering
\includegraphics[width=0.95\columnwidth]{ 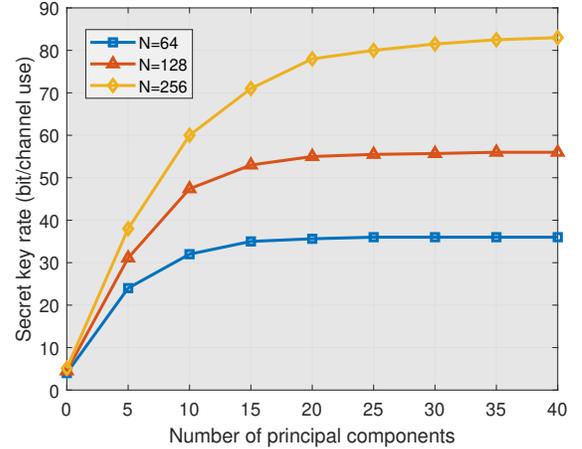}
    \caption{\footnotesize Secret key rate as a function of array size in \ac{mMIMO}.}
    \label{fig:mMIMO_SKG}
\end{figure}

5G and beyond networks include new features such as \ac{mMIMO} architectures, \ac{UWB} signaling, and migration to higher frequency bands which makes the channel highly sparse in both time and space. Unlike the rich scattering channels observed in the legacy networks, sparse channels put individual features (such as \ac{AoA}, \ac{AoD}, and delay) of the \acp{MPC} at the disposal of \ac{PLS}. That is, experts can now exploit these channel features for \ac{PLS} instead of the effective channel fading coefficients from \ac{CSI}. For instance, the study in \cite{jiao2018physical} considers \ac{AoA} and \ac{AoD} in \ac{mMIMO} as the common source of randomness between the legitimate nodes for key generation. Apart from being easy to estimate, i.e., accessibility, (through virtual channel estimation approaches), these features have been shown to be robust against noise and exhibit high reciprocity even in the low \ac{SNR} regimes. Furthermore, sparsity has made it possible to exploit the frequency-invariant features of the multipath channel, such as \acp{AoA}, \acp{AoD}, and path delays, which exhibit reciprocity even in the \ac{FDD} systems. Figure \ref{fig:mMIMO_SKG} shows a trend of secret key rate in beam space channel as a function of array size. The secret key was generated from virtual \acp{AoA}/\acp{AoD}, considering different number principle components. As the array size increases, the spatial uniqueness of the observed channel is improved and the accuracy with which the principle components are estimated improves, thereby enhancing the secret bit agreement ratio and the secrecy rate in general.

\subsection{Array Non-Stationarity}
In addition to the high spatial resolvability of the \acp{MPC}, large dimensional arrays, \ac{mMIMO} in particular, have been found to experience other peculiar channel features that can be beneficial in various aspects of wireless communication including \ac{PLS}. Essentially, different channel parameters such as gain, $K-$factor, angular power distribution, and correlation level have been found to vary greatly across array elements. Such significant variations of multiple channel parameters not only increase channel entropy for stronger keys generation but also decorrelate the channel coefficients in different domains, thereby enhancing uniqueness. Two phenomena, \ac{SW} and cluster \ac{VR}, are considered to be the main factors contributing to the peculiar channel characteristics in \ac{mMIMO}: 

\textit{Spherical Wavefront (\ac{SW}):} With large arrays, most of the significant scatters and \acp{UE} are located within the near-field region of the array in which the traditional planar wavefront assumption is no longer valid. In this case, the signal arrives at the array with \ac{SW} which has been found to add an extra phase shift that greatly enhances the spatial channel decorrelation. Some early analyses have revealed that \ac{SW} propagation is capable of decorrelating the channels of closely-located \acp{UE} even in the \ac{LoS} scenarios, thereby enhancing spatial uniqueness.  This enhances system's ability to spatially distinguish legitimate \ac{UE} from the illegitimate ones for better \ac{PLS} implementation.


\textit{Cluster \ac{VR}:} The fact that the majority of the significant scatterers fall within the near-field region of the array, not every scatterer is "visible" to each element in the array. This brings the concept of cluster \ac{VR} in which different array elements experience different channel responses based on the group of scatters visible to them. This makes the channel highly non-stationary across the array, creating additional spatial channel decorrelation at both transmitter and receiver sides which is advantageous for \ac{PLS}. The non-stationarity can be observed in different channel aspects, such as \ac{AoA}/\ac{AoD}, gain, and fading statistics across the array which can be individually or collectively exploited for \ac{PLS}.



\subsection{Temporal, Doppler and Spatial Non-Stationarity}
The non-stationarity concept also spans to temporal and Doppler domains of the channel. The non-stationarity in these domains is observed particularly in scenarios characterized with high mobility, such as those involving \ac{HST} and \ac{V2V}/\ac{V2X} communication. Unlike the conventional fixed-to-mobile cellular system, these scenarios are characterized by excessive mobility from both the terminals and the surrounding scatterers which makes the propagation conditions change rapidly. The set of scatterers interacting with the signal varies constantly with time, leading to the \textit{cluster fade-in fade-out} effect. Consequently, spatial, temporal, and Doppler characteristics of the observed \acp{MPC} vary rapidly and continuously. Such multi-domain rapid variation of the propagation characteristics can be leveraged to design cross-domains \ac{PLS} techniques. For example, a security key can be generated by harvesting channel randomness from more than one domain \cite{mazloum2015analysis}. However, as discussed before, the rapidly varying channels often lead to channel aging problems which disrupt reciprocity.

\section{Role of Channel Control and Sensing on \ac{PLS}} \label{sec:ChannelControl}
The ability to control and optimize the wireless propagation characteristics is envisioned as one of the prominent features of future networks. Such ability is discussed under the concept of smart/intelligent radio environment whose realization relies on technologies such as \ac{RIS}, \ac{ABC}, \ac{ML}, sensing, and \ac{REM}. Indeed, this ability will bring spillover benefits to different aspects of the network such as capacity improvement, interference management, security enhancement, etc. This section explores different channel control mechanisms and sensing from \ac{PLS} perspective.

\subsection{Channel Control for PLS}
Majority of the ongoing studies on channel control are steered toward \ac{RIS}-based control in which the channel is controlled from within the propagation environment. However, with the advent of the reconfigurable/smart \ac{RF} front-ends and antennas concepts, as well as the advancement in the signal processing techniques, the channel control ability can be imparted on the network at any point along the line of communication process, i.e., at the baseband processing, \ac{RF} front-end stage, antenna configuration, or within the propagation environment. As such, the discussion here highlights the potentials of different channel control at different stages for \ac{PLS}.

\subsubsection{Baseband Processing-based Channel Control}
\begin{figure}[t]
    \centering
\includegraphics[width=1\columnwidth]{ 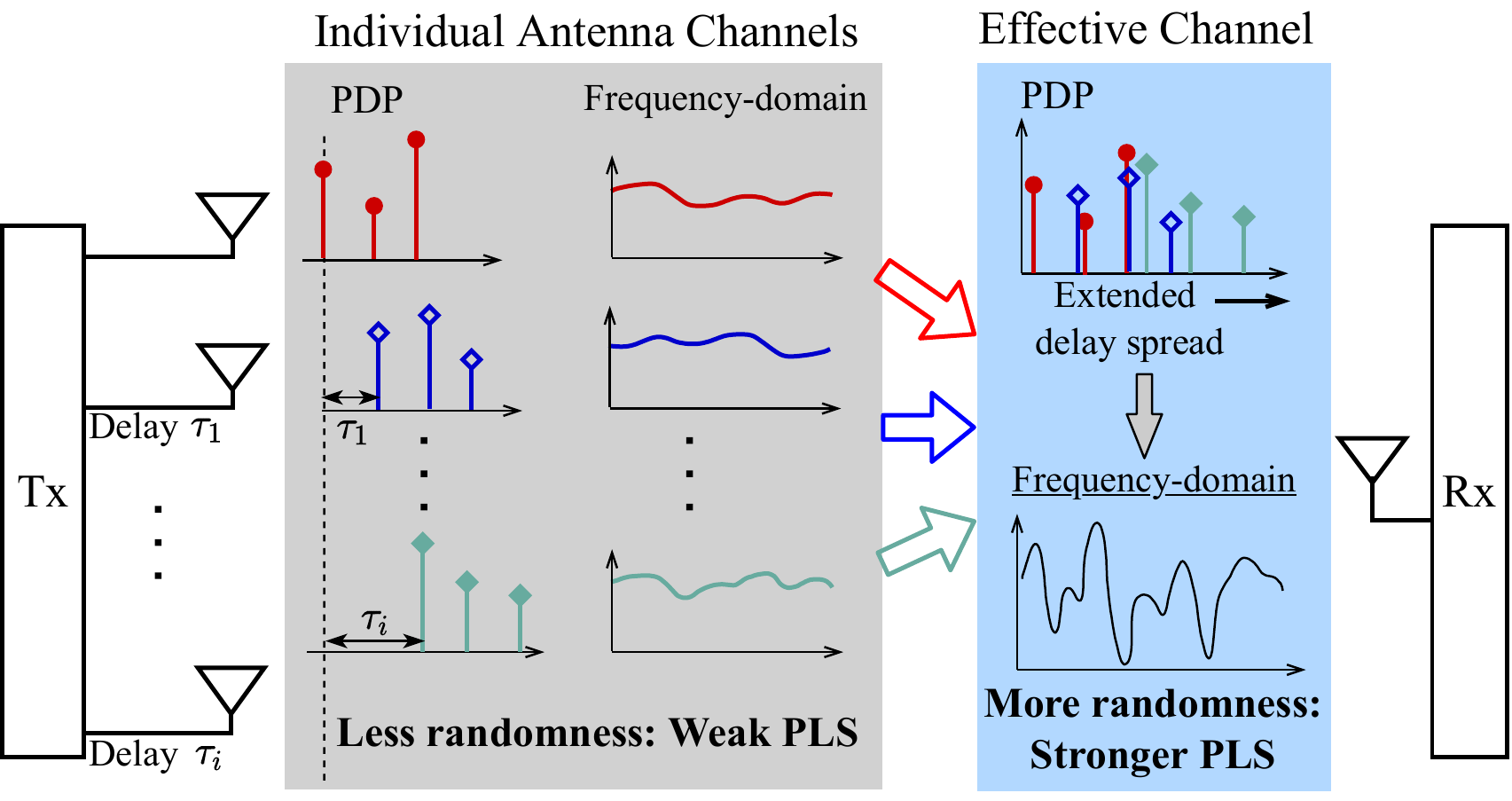}
    \caption{\footnotesize Illustration of DD for randomness enhancement for stronger \ac{PLS} implementation}
    \label{fig:DDsystem}
\end{figure}

Several digital baseband signal processing techniques such as \ac{DD}, \ac{DoD}, opportunistic beamforming, and their variants have been used to impart artificial fast fading effects, particularly in the poor scattering environment. The \ac{DD} approach, for example, artificially extends the channel's delay spread by transmitting delayed versions of the same signal from different antennas, as illustrated in Fig. \ref{fig:DDsystem}. The resultant extended delay spread corresponds to the increased selectivity in the frequency domain which can be exploited for diversity as well as security. In \cite{sun2015artificial}, \ac{CDD} technique is used to facilitate a covert \ac{OFDM} signal transmission against eavesdropping attacks. Channel shortening approach which employs a channel-specific \ac{FIR} filter to manipulate the channel's delay spread has been used to induce \ac{ISI} to the attacker's received signal and degrade its performance without harming the legitimate user \cite{furqan2017enhancing}. Essentially, different channel features can be biased through signal processing to degrade attacker's link while favoring the legitimate one, as exemplified above.

\subsubsection{Antenna Architecture-based Channel Control}
The manner in which the signal is emitted to the environment and captured by the receiving terminal has a significant impact on the effective channel response. This suggests that antenna structure plays an important role in how the environment responds to the signal traversing it. Transceiver structures utilizing \ac{mMIMO} architectures or \acp{RA}, for example, are flexible enough to manipulate their radiation pattern and effectuate some channel characteristics desirable for the ongoing communication. From the \ac{PLS} point of view, randomness, uniqueness, and even accessibility of certain channel features can be controlled through radiation pattern manipulation. For instance, the study in \cite{jiao2019physical} leverages the beamwidth and beam-direction control capability of \ac{mMIMO} to cherry-pick the environmental scatters that should interact with the transmitted signal in order to reduce temporal correlation and enhance the security key entropy rate. Likewise, \acp{RA} are capable of generating different radiation patterns that interact differently with exactly the same propagation environment, producing independent fading patterns which enhance the randomness and uniqueness of the observed channel response. This particular capability of \acp{RA} has been exploited by media-based and spatial modulation concepts to enhance spectral efficiency. \ac{SKG} and adaptation-based security approaches can leverage these capabilities as well. Unlike \ac{mMIMO}, \acp{RA} have smaller form factor that makes them suitable for imparting channel control capability in \ac{IoT} devices.

\subsubsection{RIS-based Channel Control}
The channel control mechanisms discussed above are implemented at the transceiver and their realization can be hindered by the capability of the devices. \Ac{RIS}-based control, on the other hand, allows the network to control the propagation condition from within the environment, thereby facilitating the accessibility of the favorable channel condition even for the low-end devices. Many \ac{PLS} techniques utilizing the exotic electromagnetic propagation characteristics effectuated by \acp{RIS} have been proposed. A comprehensive summary of these techniques can be found in our recently published survey \cite{almohamad2020smart}.


\subsection{Sensing for PLS}
In the future, various sensing methods such as Radar, Lidar, computer vision, and \ac{RF}-based sensing are envisioned to enrich the networks with detailed and reliable knowledge about the propagation characteristics. Particularly, sensing will help the network to acquire a variety of hidden features in the environment, such as scatterers' locations, shapes, materials, and mobility patterns, to mention just a few. Furthermore, with visual data from sensing, networks will be able to predict some events such as the occurrence of \ac{LoS} and non-\ac{LoS} propagation. All these can be exploited to adapt the transmission characteristics with respect to the legitimate nodes to enhance privacy. Our recently published survey \cite{furqan_2021_REMsec} gives more details on various ways of exploiting sensing for \ac{PLS}.

\section{Integrity of Channel Features} \label{sec: Integrity}
As \ac{PLS} becomes popular, numerous types of attacks have been launched make the channel unsuitable for \ac{PLS} application or circumvent the applied channel based \ac{PLS} algorithms. We categorize these attacks into \textit{direct} and \textit{indirect} attacks and discuss them below: 

 There are two types of integrity attacks on channel; signal processing based integrity attacks and environmental based Integrity attacks. The fırst category ıs based  on desıgnıng dıfferent sıgnal processıng technıuqe to extract channel state ınformatıon from precodıng matrıx or fılterıng based technıques whıle the later technıques are based on dırect manıpulatıon of wırless envıronment by RIS or 

 Although it is possible that individual channel features are eligible to enhance PLS, these features can be manipulated by the attacker to degrade the performance of different PLS algorithms. In this section, some of the attacker's techniques are investigated from the perspective of individual channel parameters. 
 \subsection{Indirect Attacks on the Integrity} \label{sec: indirectAttacks}
In these case the adversaries implement some signal processing techniques to manipulate certain features of the channel and deceive the legitimate nodes during the channel probing stage or authentication process. Additionally, machine learning, blind signal analysis, or reverse-engineering processing of the received signal can also be applied to derive the security parameters extracted from the channel by legitimate nodes. Specific examples are given below.

\subsubsection{\Ac{CSI} Inference Attack:} 
\textit{\Ac{CSI} Inference Attack:} As mentioned earlier, most of the traditional \ac{PLS} techniques are based on utilizing \ac{CSI}, \ac{CIR} inprticular, under the spatial decorrelation assumption. However, it has been shown that, in the practical scenarios, the attacker can make use of the common knowledge about the nature of the communication scenario and some signal processing techniques to obtain the channel response of the legitimate nodes even under spatial decorrelation effect. For example, in the \textit{\ac{CSI}-snoop}\cite{zhang2018csisnoop} , the attacker exploit the knowledge of the standard/known frame structure in multi-user \ac{MIMO} networks and filtering mechanisms to compute the \ac{CSI}. Likewise, in the \textit{predictable channel attacks}, knowledge of the propagation environment geometry is utilized to precompute the \ac{CIR} between legitimate nodes. In some other cases, the attacker can take advantage of the actuality of the propagation condition such as sparse scattering, presence of a strong \ac{LoS} component or waveguide propagation effect in which the half-wavelength spatial decorrelation assumption does not hold to obtain the \ac{CSI}. These kind of attacks urge reconsideration of the \ac{PLS} approaches that rely on the effective \ac{CSI} such as \ac{CIR}. Stronger approaches can be devised by exploiting the novel, distinct channel features explored above. Similar to the cross-layer security concept in the which the PHY and MAC layer characteristics of transmission are jointly utilized, these distinct channel features in different domains can also be utilized simultaneously to increase the entropy of security keys.

\subsubsection{Blind signal analysis against noise-based}
Artificial Noise (AN)-aided \ac{PLS} are very effective against eavesdropping attacks. However, it is reported in \cite{wen2020robust} that an attacker can use blind source separation (BSS) and independent component analysis (ICA) to separate the Gaussian noise from the information signal. This urges reconsideration of the design of AN signal that is robust to such attacks.

\begin{figure}[t]
   \centering
\includegraphics[width=1\columnwidth]{ 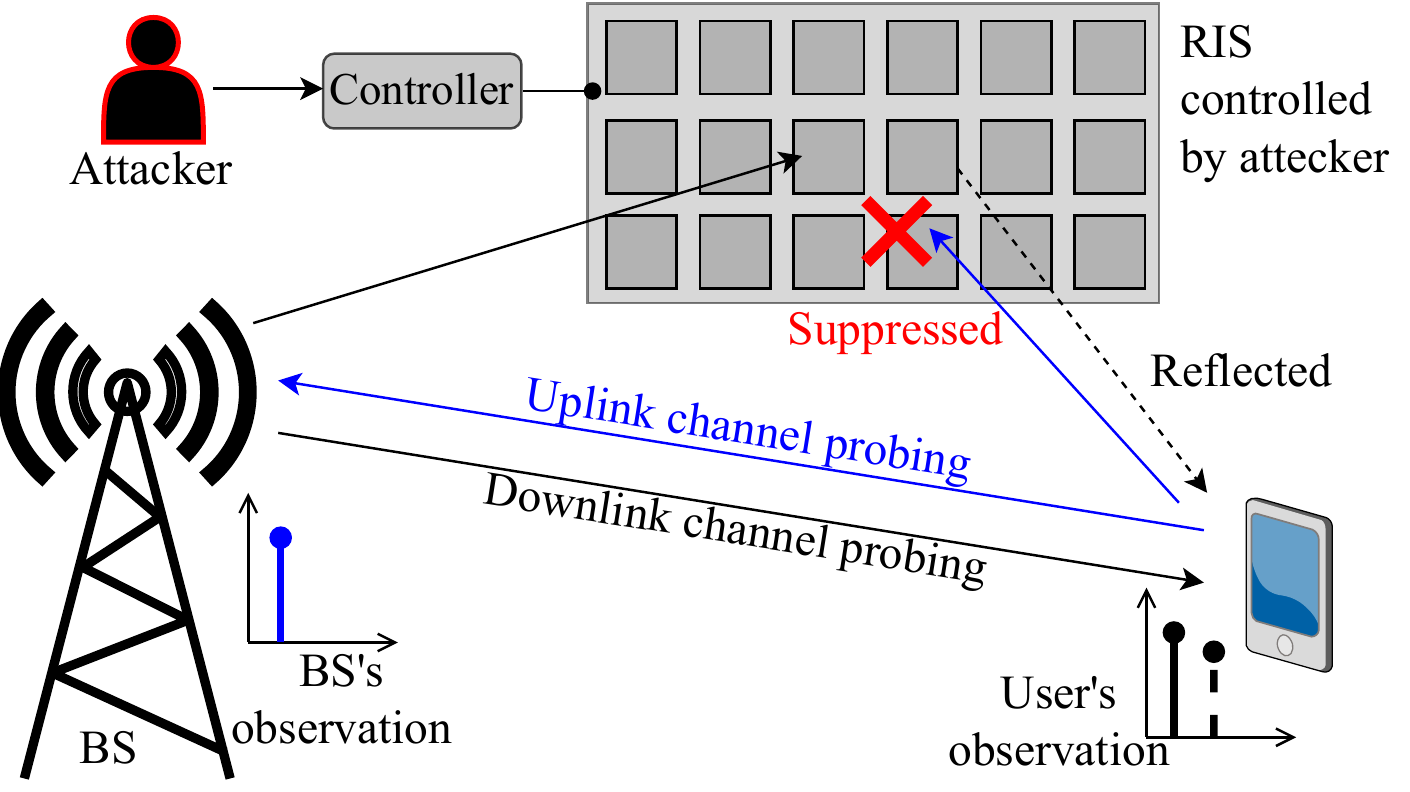}
    \caption{\footnotesize An attacker using RIS to manipulate \acp{MPC} during channel probing stage and destroy reciprocity.}
    \label{fig:RIS reciprocity attack}
\end{figure}

\subsection{Direct Attacks}
These attacks involve direct manipulation of the channel features observed by the legitimate nodes. In this case, the legitimate nodes may rely on the wrong channel characteristics (i.e., artificially generated by the attacker) to devise a \ac{PLS} algorithm or perform authentication procedures. \textit{Camouflage attack} \cite{fang2014you} is a typical example. In this case, a malicious transmitter can manipulate his own channel through some channel control mechanisms discussed in Section \ref{sec:ChannelControl} to hide his location or mimic some movement to deceive the legitimate receiver. An attacker can also use \ac{RIS} to manipulate multipath characteristics of the legitimate link during the channel probing stage using legitimate nodes' probing signals and destroy different channel characteristics, as exemplified in Fig. \ref{fig:RIS reciprocity attack}. Exploiting \ac{MoS} characteristics in \ac{THz} frequencies, the attacker can create an artificial gaseous cloud to scatter the signal to its own direction and improve its listening capability, or just degrade reliability at the legitimate node as in the case of the jamming attack. The direct attacks on the environment can also be dedicated to the \ac{PLS} approaches utilizing propagation characteristics obtained through sensing. \textcolor{black}{For example, the attacker can rebroadcast the sensing signal with some modification or use \ac{RIS} to manipulate the speed, range, or trajectory of the environmental objects or cause total misdetection of these objects, leading to the acquisition of wrong surrounding information.}   

\section{Future Direction}
\subsection{Securing Integrity of the Channel Features}
As discussed in Section \ref{sec: Integrity}, a number of attacks dedicated directly to the propagation environment characteristics have emerged, which can potentially jeopardize the credibility of the channel-based \ac{PLS}. This calls for the widening of the \ac{PLS} scope from securing the communication process to securing the propagation environment as well. In fact, security of the propagation characteristics is not critical for the \ac{PLS} alone but for other sensitive applications, such as \ac{REM}, as well. This opens up a new research frontier for the \ac{PLS} concept.

\subsection{High Mobility and Non-Stationarity Issues} 
High mobility experienced in vehicular communication makes channels extremely random in all domains, which is desirable for \ac{PLS} applications. However, the rapidly varying channel is accompanied by the channel aging problem that may disrupt channel reciprocity as discussed before. Furthermore, for mobile users, beam tracking is necessary. Therefore, more effective \ac{PLS} algorithms that consider the effects of different mobility-related issues need to be devised.

\subsection{Beam Squint Issue in \ac{mMIMO}}
The use of \ac{UWB} signaling on \ac{mMIMO} transceivers causes the spatial-wideband effect or beam-squint problem \cite{wang2018spatial_2}. Beam-squint causes beam spreading in multi-carrier systems which affects \acp{MPC}' \acp{AoD} and \acp{AoA}. This destroys the reciprocity property of these features, rendering them inappropriate for \ac{PLS}. Impact of beam squint effect on the \ac{PLS} approaches utilizing \acp{AoD} and \acp{AoA} in \ac{mMIMO} systems is an open problem.

In the antenna array systems, the received signal at different array elements is a slightly delayed version of the original signal. With large array dimensions such as \ac{mMIMO}, substantial amount of delay can be observed across the array elements. In case \ac{UWB} signaling (short symbol duration) is used, the delay may become comparable or even larger than the symbol duration, leading to the so-called delay squint problem which translates into beam squint problem in the spatial domain. Beam squint affects the \acp{MPC}' \acp{AoD} and \acp{AoA}. This destroys the reciprocity property of these features, rendering them inappropriate for \ac{PLS}. Impact of beam squint effect on the \ac{PLS} approaches utilizing \acp{AoD} and \acp{AoA} in \ac{mMIMO} systems is an open problem.

\subsection{Intelligent Security Frameworks}
  The increasing flavors of wireless networks, varying capabilities of devices, and different security requirements of users and applications motivate the need for adaptive, intelligent, and flexible security designs. The intelligent and flexible design can be designed by exploiting cross-layer coordinated security architecture. based on joint information from physical, MAC, network, and application layers. The  effect of individual channel parameters on \ac{PLS} will help in designing efficient cross-layer intelligent security frameworks \cite{yilmaz2017cognitive}.

\section{Conclusion}
This work revisits the channel-based \ac{PLS} concept considering the newly acquired network capabilities in accessing and controlling different channel characteristics. It thus draws attention to the importance of individual channel features in enabling efficient \ac{PLS} techniques. In this regard, various opportunities are discussed and challenges are highlighted to ignite further research. It is however important to emphasize that some of these features are still being studied and only their tentative models are available. Therefore, the discussion herein only intends to reveal their potential from the \ac{PLS} point of view and encourage further research in that direction.

\begin{IEEEbiographynophoto}{Abuu B. Kihero} received his B.S. and M.S. in electronics engineering from Gebze Technical University and Istanbul Medipol University, respectively. He is currently pursuing his Ph.D. degree in Istanbul Medipol University where he is also working as a graduate researcher. His research focuses on wireless channel characterization and modeling, cooperative networks, and waveform design.

\end{IEEEbiographynophoto}
\begin{IEEEbiographynophoto}{Haji M. Furqan} received his B.S. and M.S. degrees in electrical engineering from COMSATS Institute of Information Technology (CIIT), Islamabad, Pakistan in 2012 and 2014, respectively. He received his Ph.D. degree from Istanbul Medipol University, Turkey, where he is currently a Post-Doc researcher. His research focuses on physical layer security, cooperative communication, adaptive index modulation, OFDM, V2X, 5G systems, and wireless channel modeling and characterization.
\end{IEEEbiographynophoto}

\begin{IEEEbiographynophoto}{Mehmet Mert {\c{S}}ahin} [S19] received the B.S. degree from Bilkent University, Ankara, Turkey, in 2019. He worked at Aselsan Inc. as a wireless communication design engineer in 2019. He is currently working toward the PhD degree at University of South Florida, Tampa, FL, USA. His research interest include waveform design, multiple-accessing, joint radar-sensing and communication.
\end{IEEEbiographynophoto}

\begin{IEEEbiographynophoto}{H{\"u}seyin Arslan} [S’95, M’98, SM’04, F’16] received his B.S. degree  in  electrical  and  electronics  engineering  from  Middle East Technical University in 1992, and his M.S. and Ph.D. degrees in electrical engineering from Southern Methodist Uni-versity, Dallas, Texas, in 1994 and 1998, respectively. He is a professor of electrical engineering at the University of South Florida and the Dean of the School of Engineering and Natural Sciences at Istanbul Medipol University, Turkey.
\end{IEEEbiographynophoto}



\end{document}